\documentclass{aa}
\usepackage{graphicx}
\usepackage{rotating,times,pictex,graphicx,latexsym}
\usepackage{color}
\begin{document}
\title{A Galactic Plane Relative Extinction Map from 2MASS}

\author{Dirk Froebrich$^1$, Thomas P. Ray$^1$, Gareth C. Murphy$^1$, \and
Alexander Scholz$^2$} 

\offprints{df@cp.dias.ie}

\institute{$^1$ Dublin Institute for Advanced Studies, 5 Merrion Square, Dublin
2, Ireland \\ $^2$ University of Toronto, Dept. Astronomy \& Astrophysics, 66
St. George's St., Toronto, Canada}

\authorrunning{D.~Froebrich et al.}

\date{Received sooner ; accepted later}

\abstract{We present three 14400 square degree relative extinction maps of the 
Galactic Plane ($|b|$\,$<$\,20\degr) obtained from 2MASS using  accumulative
star counts (Wolf diagrams). This method is independent of the colour of the
stars and the variation of extinction with wavelength. Stars  were counted in
3\arcmin$\!\!$.5\,$\times$\,3\arcmin$\!\!$.5 boxes, every 20\arcsec. 
1\degr$\times$1\degr\, surrounding fields were chosen for reference, hence the
maps represent local extinction enhancements and ignore any contribution from
the ISM or very large clouds. Data reduction was performed on a Beowulf-type
cluster (in approximately 120 hours). Such a cluster is ideal for this type of
work as areas of the sky can be independently processed in parallel. We
studied  how extinction depends on wavelength in all of the high extinction
regions detected and within selected dark clouds. On average a power law
opacity index ($\beta$) of 1.0 to 1.8 in the NIR was deduced. The index however
differed significantly from region to region and even within individual dark
clouds. That said, generally it was found to be constant, or to increase, with
wavelength within a particular region.  

}

\maketitle
%

\section{Introduction}

Dust is not only one of the most important coolants in molecular clouds but is
also an excellent tracer of molecular hydrogen. Determining its distribution is
thus crucial to understand the early stages of star formation. It gives us
information, for example, on how clouds fragment and ultimately this knowledge
must be reconciled with the initial mass function  (Padoan, Nordlund, \& Jones
\cite{1997MNRAS.288..145P}). Moreover by combining dust extinction maps with
infrared and submillimeter dust emission maps, we can derive basic knowledge of
grain properties and search for evidence of grain evolution.  

Extinction towards an individual star can of course by determined by measuring
its colour excess although this method requires knowledge of the star's
intrinsic colour (e.g. He et al. \cite{1995ApJS..101..335H}, Racca et al.
\cite{2002AJ....124.2178R}). The technique is also highly selective as it only
gives us information along very precise lines of sight. An alternative
approach, developed many years ago (Wolf \cite{1923AN....219..109W}, Bok 
\cite{1956AJ.....61..309B}), and particularly appropriate for dark clouds, is 
to count the number of stars in the cloud's vicinity down to a limiting 
magnitude and compare this with a control ``unextincted'' region nearby. Such
so-called Wolf Diagrams are of limited use in the optical, as one can only
probe the outer peripheries of a cloud where the extinction is low. The method
works well however in the infrared, due to its improved transmission with
respect to the optical, allowing regions with A$_{\rm V}$ values as high as
30--40 to be measured. 

Dust not only absorbs light from background stars but also reddens it. By 
measuring extinction in a number of wavebands, one can determine how it varies
with wavelength ($\lambda$). Normally this variation is assumed to be a power
law, i.e. A$_\lambda \propto \lambda^{-\beta}$  where $\beta$ is the opacity
index. A constant value for $\beta$ however is not necessarily guaranteed {\em
a priori}.  

One drawback in the past of the star counting method, at least for large areas,
is that it is computationally very expensive. This type of data reduction,
however, is  ideally suited to a parallel computing environment provided, for
example, by a Beowulf-type cluster. Large swathes of the sky can be processed
simultaneously  and independently of each other. Moreover, the possibility for
exploiting such ``number crunching'' facilities coincides with the availability
of large ground based IR surveys such as 2MASS and DENIS. For the first time
its possible to manufacture extinction maps covering large fractions of the
sky. 

Here we present a high spatial resolution relative extinction map of the
Galactic Plane, in J, H, and K, derived from the 2MASS database using the star
count method. The method and its limitations are described in detail in
Sect.\,2, the determination of the noise is discussed in Sect.\,3. Results of
the opacity index computations are presented in Sect.\,4 and our conclusions
are given in Sect.\,5.     

\begin{figure*}[t]
\beginpicture
\setcoordinatesystem units <0.45mm,0.45mm> point at 0 0
\setplotarea x from -180 to 180 , y from -20 to 120
\put {\includegraphics[width=16.24cm,height=1.845cm, bb=14 14 375 55]{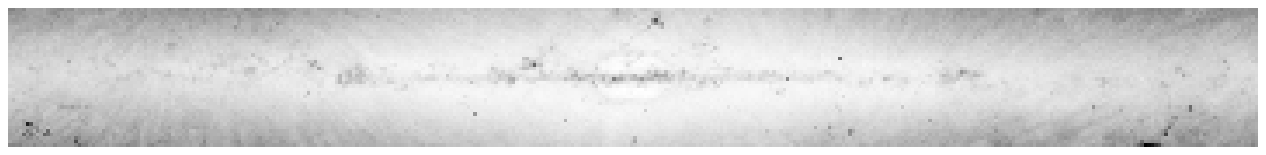}} at 0.2 0.2
\put {\includegraphics[angle=90,width=16.18cm,height=1.78cm]{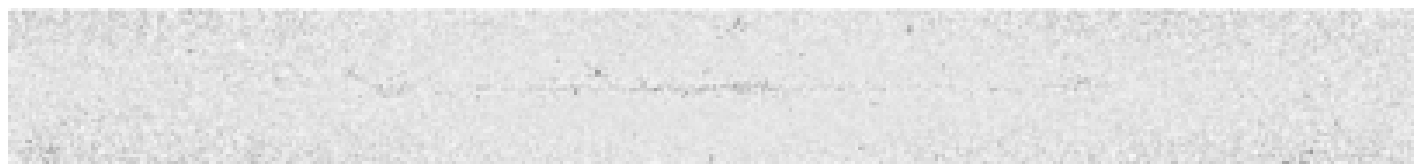}} at 0 50.5
\put {\includegraphics[width=16.2cm,height=1.8cm]{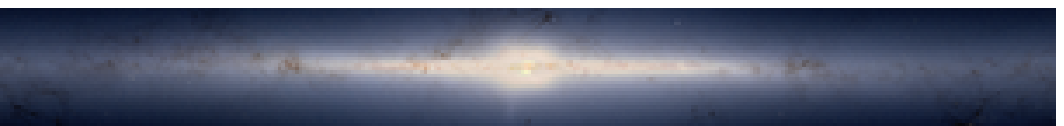}} at 0 100
\axis left label {$b$\,[$^\circ$]\,\,\,\,\,\,\,\,\,\,}
ticks in long unlabeled from -20 to 120 by 10
      short unlabeled from -20 to 120 by 10 /
\axis right label {}
ticks in long unlabeled from -20 to 120 by 10
      short unlabeled from -20 to 120 by 10 /
\axis top label {}
ticks in long unlabeled from -180 to 180 by 60
      short unlabeled from -180 to 180 by 20 /
\put {180} at -180 -26
\put {120} at -120 -26
\put {60} at -60 -26
\put {0} at 0 -26
\put {$l$\,[$^\circ$]} at 0 -35
\put {300} at 60 -26
\put {240} at 120 -26
\put {180} at 180 -26
\put {$-$20} at -190 -20
\put {\,\,\,\,\,\,\,\,0} at -190 0
\put {$+$20} at -190 20
\put {$-$20} at -190 30
\put {\,\,\,\,\,\,\,\,0} at -190 50
\put {$+$20} at -190 70
\put {$-$20} at -190 80
\put {\,\,\,\,\,\,\,\,0} at -190 100
\put {$+$20} at -190 120
\plot 57 40 60 40 60 44 57 44 57 40 /
\setplotarea x from -180 to 180 , y from -20 to 20
\axis top label {}
ticks in long unlabeled from -180 to 180 by 60
      short unlabeled from -180 to 180 by 20 /
\axis bottom label {}
ticks in long unlabeled from -180 to 180 by 60
      short unlabeled from -180 to 180 by 20 /
\setplotarea x from -180 to 180 , y from 30 to 70
\axis top label {}
ticks in long unlabeled from -180 to 180 by 60
      short unlabeled from -180 to 180 by 20 /
\axis bottom label {}
ticks in long unlabeled from -180 to 180 by 60
      short unlabeled from -180 to 180 by 20 /
\setplotarea x from -180 to 180 , y from 80 to 120
\axis bottom label {}
ticks in long unlabeled from -180 to 180 by 60
      short unlabeled from -180 to 180 by 20 /
\endpicture
\caption{\label{noise} JHK colour composite of the star count map (top),
relative extinction map obtained from the J-band data (middle) and three sigma
noise due to the non-uniform distribution of stars in the relative extinction
maps for J (bottom). The noise is displayed in linear scale from zero (white)
to 0.7\,mag (black) of optical extinction. The rectangle marks the region
magnified in Fig.\,\ref{musca}.}  
\end{figure*}

\section{Method and Limitations} 
\label{limits}

Relative extinction maps were determined using Wolf diagrams (e.g. Kiss et al.
\cite{2000A&A...363..755K}). At each position accumulated star counts were
performed in a small box. These counts were compared to the average accumulated
star count in a (co-centred but larger) comparison field to determine the
relative extinction. The method is independent of stellar colour and variation
of extinction with wavelength. A number of assumptions, however, have to be
made (e.g. Froebrich et al. \cite{2005A&A.in.press.F}): (1) stars are
distributed uniformly and all apparent voids are due to extinction (2) one may
define a constant average absolute magnitude for the stars in a box (3) the
completeness limit of the catalogue does not depend on position. In reality
these assumptions may only be valid under certain conditions. For example
reference to an average absolute magnitude is only meaningful if the box
contains enough stars. Moreover stars are not distributed uniformly on the sky
but if the control field is not too far away, the assumption of uniformity is
locally valid. Thus, to apply the method properly one must ensure that the
choice of box size, control field, etc., are within certain limits (see below).

Using the 2MASS catalogue already ensures a high level of uniformity and
quality of the photometry. To avoid photometric errors for faint objects we
only selected sources with a signal to noise (S/N) ratio $>$\,5. This
corresponds to quality flags A/B/C in the 2MASS catalogue, and an error in
stellar magnitude of at most 0.17\,mag. Such a value is slightly higher than
our stepsize in brightness (counts were performed every 0.1\,mag). However, the
extinction is derived using an average for all stars in the box and is thus
determined with greater precision.  

\begin{figure*}[t]
\beginpicture
\setcoordinatesystem units <16.5mm,16.5mm> point at 0 0
\setplotarea x from 0 to 10 , y from -10 to -6
\put {\includegraphics[angle=0,width=4.75cm,height=6.38cm, bb=30 16 183 128]{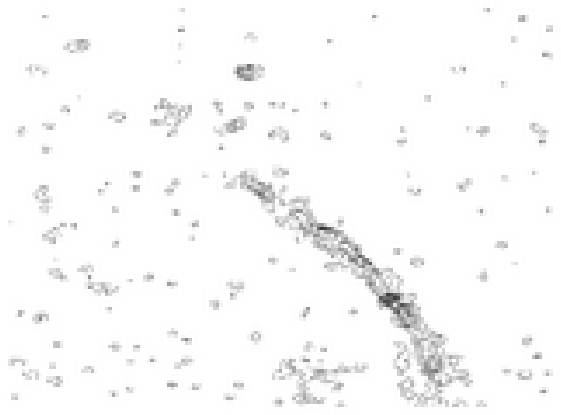}} at 1.5 -8
\put {\includegraphics[angle=0,width=4.75cm,height=6.38cm, bb=4 4 175 234]{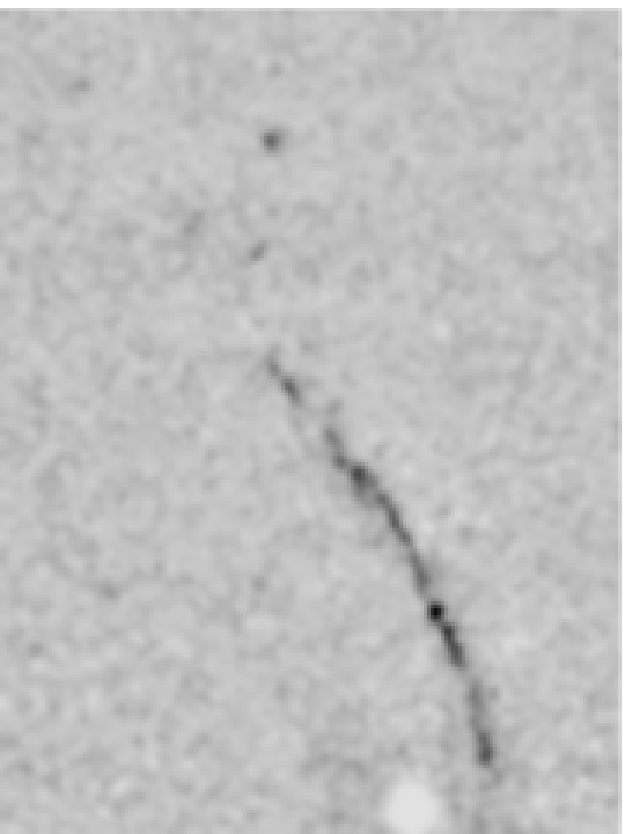}} at 5.0 -8
\put {\includegraphics[angle=0,width=4.75cm,height=6.38cm, bb=2 3 115 156]{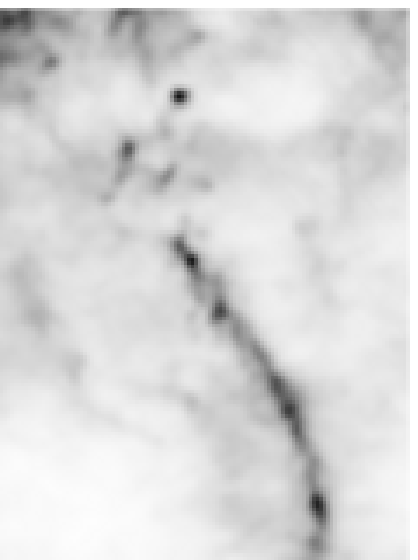}} at 8.5 -8
\put {-10} at -0.2 -10
\put {- 9} at -0.2 -9
\put {- 8} at -0.2 -8
\put {- 7} at -0.2 -7
\put {- 6} at -0.2 -6
\put {303} at 0 -10.2
\put {302} at 1 -10.2
\put {301} at 2 -10.2
\put {300} at 3 -10.2
\put {303} at 3.5 -10.2
\put {302} at 4.5 -10.2
\put {301} at 5.5 -10.2
\put {300} at 6.5 -10.2
\put {303} at 7 -10.2
\put {302} at 8 -10.2
\put {301} at 9 -10.2
\put {300} at 9.85 -10.2
\put {$l$\,[$^\circ$]} at 5 -10.4
\plot 1.4 -10 1.4 -9.4 
     2.6 -9.4 2.6 -10 /
\plot 4.9 -10 4.9 -9.4 
     6.1 -9.4 6.1 -10 /
\put {\tiny NGC} at 2 -9.7
\put {\tiny 4372} at 2 -9.83
\put {\tiny\bf *} at 1.42 -6.3
\put {\tiny\bf *} at 0.3 -9.1
\put {\tiny\bf *} at 0.65 -8.85
\put {\tiny\bf *} at 1.35 -8.8
\put {\tiny \object{S\,152}} at 0.6 -6.4
\put {\tiny \object{BHR\,79}} at 1.65 -6.6
\put {\tiny \object{BHR\,82}} at 0.9 -6.85
\put {\tiny \object{BHR\,80}} at 1.2 -7.35
\put {\tiny 14} at 1.6 -7.15
\put {\tiny DCld} at 1.7 -7.6
\put {\tiny ?} at 1.7 -7.85
\put {\tiny ?} at 1.7 -8.6
\put {\tiny 15} at 1.2 -7.75
\put {\tiny 13} at 1.3 -7.9
\put {\tiny 12} at 1.45 -8.05
\put {\tiny 11} at 1.5 -8.2
\put {\tiny 10} at 1.85 -8.15
\put {\tiny 9} at 1.9 -8.25
\put {\tiny 8} at 1.95 -8.35
\put {\tiny 7} at 2 -8.45
\put {\tiny 6} at 1.93 -8.75
\put {\tiny 5} at 1.98 -8.92
\put {\tiny 4} at 2.05 -9.12
\put {\tiny 3} at 2.2 -9.3
\put {\tiny 2} at 2.5 -9.3
\put {\tiny 1} at 2.5 -9.6

\axis left label {$b$\,[$^\circ$]\,\,\,\,\,\,\,\,\,\,}
ticks in long unlabeled from -10 to -6 by 1
      short unlabeled from -10 to -6 by 0.2 /
\axis right label {}
ticks in long unlabeled from -10 to -6 by 1
      short unlabeled from -10 to -6 by 0.2 /
\axis bottom label {}
ticks in long unlabeled from 0 to 10 by 0.5
      short unlabeled from 0 to 10 by 0.5 /
\axis top label {}
ticks in long unlabeled from 0 to 10 by 0.5
      short unlabeled from 0 to 10 by 0.5 /
\setplotarea x from 0 to 3 , y from -10 to -6
\axis right label {}
ticks in long unlabeled from -10 to -6 by 1
      short unlabeled from -10 to -6 by 0.2 /
\setplotarea x from 3.5 to 6.5 , y from -10 to -6
\axis right label {}
ticks in long unlabeled from -10 to -6 by 1
      short unlabeled from -10 to -6 by 0.2 /
\axis left label {}
ticks in long unlabeled from -10 to -6 by 1
      short unlabeled from -10 to -6 by 0.2 /
\setplotarea x from 7 to 10 , y from -10 to -6
\axis left label {}
ticks in long unlabeled from -10 to -6 by 1
      short unlabeled from -10 to -6 by 0.2 /
\endpicture
\caption{\label{musca} Relative extinction map obtained from J-band 2MASS data
of the Musca Dark Cloud. Contours in the left panel start at A$_{\rm
V}$\,=\,1\,mag and are in steps of 1\,mag. We marked previously known and new
regions with peak extinction values larger than A$_{\rm V}$\,=\,2\,mag.
``Fake'' globules due to bright stars are labelled with *, and numbers
represent regions from Vilas-Boas et al. (\cite{1994ApJ...433...96V}) (MU\,16
could not be detected). The middle panel shows the relative extinction map in
grey-scale and the right panel the 100\,$\mu$m IRAS image of this region. In
the marked rectangle, around \object{NGC\,4372}, extinction values are less
reliable due to additional noise (see also Sect.\,\ref{limits}).}  
\end{figure*}

As we have emphasised, accumulated star counts are only meaningful, providing
boxes contain a sufficient number of stars. In particular the S/N in the final
relative extinction maps strongly depends on the number of stars in the area
being counted and hence on the box-size of the reseau (see Appendix C in
Froebrich et al. \cite{2005A&A.in.press.F} for a detailed discussion). Taking
into account the density of stars in the 2MASS catalogue in all 3 filters, we
selected a box-size of 3\arcmin$\!\!$.5\,$\times$\,3\arcmin$\!\!$.5 to perform
the counts. This ensures approximately 25 stars per box (S/N\,=\,5, assuming a
Poisson distribution) at the completeness limit for 99.5, 73.5, 63.5\,\% of the
area with $|b|$\,$<$\,10\degr, in J, H, and K, respectively. It also leads to a
mean density of about 100 stars or more per box in the J-band for a large area
($\approx$\,2800\,$\Box\degr$). Such a density converts to around one star per
20\arcsec$\times$20\arcsec. Hence performing star counts with a spatial
frequency of 20\arcsec\, ensures optimal sampling of the 2MASS catalogue. Note
that this oversampling of about 10 is not unconditionally required in regions
with fewer stars, but was kept to ensure a uniform pixel size throughout the
whole extinction map.

Counts in the 3\arcmin$\!\!$.5\,$\times$\,3\arcmin$\!\!$.5 box where compared
with a co-centred larger control field. For the latter we chose a
1\degr$\times$1\degr\, area. This has the effect that our maps record local
extinction enhancements, due to small clouds or globules, and not global
extinction effects due to the ISM or extended dark clouds.

Achieving a high S/N in our maps depends on extending the star counts to the
faintest possible magnitude (and thus the largest possible number of stars).
Care however has to be taken as the completeness limit of the catalogue varies
strongly with galactic coordinates. We obtained this limit for each position by
determining the peak in the histogram of stars per magnitude bin in the 
associated control field. Our relative extinction maps are then obtained
exclusively using stars 0.3\,mag brighter than this completeness limit.

The exercise of counting stars can be described as an embarrassingly parallel 
problem, as results from one part of the sky do not depend on any other. The
entire 2MASS catalogue covering $|b|$\,$<$\,20\degr\, from the Galactic Plane 
was processed with a sampling frequency (effective pixel size) of 20\arcsec\,
using our 32-node Beowulf-type cluster of 2.67\,GHz P4 HyperThreat processors.
The three 14400$\Box$\degr\, relative extinction maps (in J, H and K) took
about 40\,hours each to calculate. 

In Fig.\,\ref{noise} we present a three colour composite (J, H and K) full size
star count map (top panel) and the relative extinction map (middle panel)
obtained from the 2MASS J-band data. The small rectangle in the figure marks
the region blown-up in Fig.\,\ref{musca}. The latter figure, of the Musca Dark
Cloud, demonstrates the full resolution obtained in our relative extinction
maps (in contours in the left and grey-scale in the middle panel). Known dark
clouds are labelled, as well as ``fake'' globules (see below). The panel to the
right of Fig.\,\ref{musca} shows for comparison the IRAS 100\,$\mu$m image and
nicely demonstrates the correspondence between dust in emission and absorption.

Comparing accumulated star counts in a
3\arcmin$\!\!$.5\,$\times$\,3\arcmin$\!\!$.5 box with the surrounding
1$^\circ$\,$\times$\,1$^\circ$ field is ideal for finding small regions of
enhanced extinction. All fields containing high extinction regions, which are
more extended than a significant portion of one square degree, however, will
not have the correct extinction values. In particular this includes the large
Orion, Taurus, and Ophiuchus clouds. Here a larger comparison field is needed,
but see the discussion in Sect.\,\ref{beta_cal}. 

Very rich star clusters (mainly globular clusters) are problematic in the sense
that star counts are anomalously  high at their centre compared to their
peripheries. This gives rise to regions of apparent negative extinction at the
cluster centre and apparent positive extinction further out. 
Fig.\,\ref{musca}, showing the globular cluster NGC\,4372, illustrates this
effect. Fortunately these regions are usually not very close to areas of real
higher extinction.  Another effect is that extremely bright stars prevent
detection of sources close by and hence also mimic  high extinction regions
(also see Fig.\,\ref{musca} for a few examples). Finally there are physical and
effective gaps in the 2MASS catalogue which also result in higher apparent
extinction. Fortunately these regions cover only 0.006\,\% of the whole sky.

\section{Noise Determination}

We determined the noise in our images to compare our method with the NICER
technique of Lombardi \& Alves (\cite{2001A&A...377.1023L}). Our relative
extinction maps show noise due to the non-uniform distribution of stars,
$\sigma_1$, and the box-size within which we count stars (so-called dither
noise), $\sigma_2$. The latter manifests itself as structures of around the
size of the box used for star-counts. We thus smoothed the original maps with a
filter of width the box-size and subtracted the resulting image from the
original to obtain the noise due to the non-uniform distribution of stars.
These noise images were then gaussian filtered (5\arcmin\, FWHM), in order to
obtain the same resolution as in Lombardi \& Alves
(\cite{2001A&A...377.1023L}), and the 3$\sigma$ noise level was determined. The
noise is transformed into A$_{\rm V}$ using the conversion factors 4.04, 6.47,
9.75 for J, H, and K, respectively, given in Mathis
(\cite{1990eism.conf...63M}). The bottom panel in Fig.\,\ref{noise} shows in
grayscale the resultant $\sigma_1$ noise in our J-band map, averaged over
1$\degr$\,$\times$\,1$\degr$. 

This noise ranges from 0.2-1.0\,mag (J), 0.4-1.8\,mag (H), and 0.6-3.8\,mag (K)
in equivalent optical extinction. The values for the J-band data are slightly
higher than obtained by the NICER technique (compared with the maps of Lombardi
\& Alves (\cite{2001A&A...377.1023L}) in the Orion region at about 
$l$\,=\,210$^\circ$, $b$\,=\,$-18^\circ$). All three noise maps show the same
principle structure with the lowest noise levels obtained north and south of
the Galactic Plane within $\pm$\,60$\degr$ longitude from the Galactic Centre.
Regions directly in the Galactic Plane and within the large dark clouds
(Ophiuchus, Orion, Taurus) are affected by higher noise, due to extended
extinction and hence a smaller number of stars. 

The additional noise, due to dithering, is only slightly dependent on the
chosen box-size. A larger number of stars in the box reduces $\sigma_2$. By
varying the box-size the value for this noise can be estimated. For our
box-size we obtain typical values of about 0.7, 1.0, 1.3\,mag optical
extinction for the J, H, K-band dither-noise, respectively. See the Appendix of
Froebrich et al. (\cite{2005A&A.in.press.F}) for a detailed description of this
effect in the context of a particular region (IC\,1396). The actual detection
limit in our relative extinction map can be determined by $\sigma_{\rm
det}$\,=\,$\sqrt{\sigma_{1}^2 + \sigma_{2}^2}$ for each position and filter
individually.

\section{NIR Opacity Index}
\label{beta_cal}

Using our three large scale maps we can determine how extinction varies in the
NIR. The power law index $\beta$ of the wavelength dependence of the extinction
can be determined using:
\begin{equation} 
\beta_{\lambda_1 \lambda_2} = \log ({\rm A}_{\lambda_1} / {\rm A}_{\lambda_2})
/ \log (\lambda_2 / \lambda_1) 
\end{equation}
where A$_{\lambda_1}$ is the extinction at a wavelength $\lambda_1$. We chose
the wavelength for the 2MASS
filters\footnote{http://www.ipac.caltech.edu/2mass/releases/allsky/doc/explsup.html}
of 1.235, 1.662, and 2.159\,$\mu$m for J, H, and K, respectively. Since we have
three filters there are three possible combinations to determine $\beta$ (JH,
JK, HK), two of which are independent. 

Care has to be taken in deriving $\beta$, since it is very sensitive to small
errors in extinction especially at low extinction values. Hence only regions
which are three sigma above the local noise level were considered. A further
more complex source of error is the size and position of our control field. Due
to extinction, a fraction $f$ of the control field might just show a fraction
$e$ of stars, compared to unextincted regions. This leads to a mean extinction
A$_{\rm C}$ in the control field and hence a smaller apparent measured
extinction A$_\lambda$\,=\,A$_{\rm real}$$-$A$_{\rm C}$. It can be shown easily
that under the assumptions $f$\,$\cdot$\,(1$-$$e$)\,$\ll$\,1 and A$_{\rm
C}$\,$\ll$\,A$_{\rm real}$, the derived $\beta$ value is not influenced, even
if the measured extinction values are wrong by $A_{\rm C}$. The size of A$_{\rm
C}$ is smaller than the noise in our maps as long as
$\sqrt{N}$\,$\cdot$\,$f$\,$\cdot$\,(1$-$$e$)\,$\le$\,1, were $N$ is the number
of stars in our small box. According to our box-size and the star density, this
holds for most regions as long as $f$ is smaller than 0.3. Considering this we
also excluded all regions where more than 30\,\% of the control field is
influenced by extinction larger than the three sigma noise from the opacity
index determination.

The distribution of $\beta_{\rm JH}$, determined over the whole area of our map
is rather broad and peaks between 1.0 and 1.8. The same wide distribution is
found for $\beta_{\rm HK}$, but at slightly larger $\beta$ values (shifted by
$+$\,0.1). The reason for this particularly broad distribution can be found
when we look at groups of dark clouds in detail. We find they can be
categorised according to whether the opacity index 1) is roughly constant, 2)
increases, or 3) decreases with wavelength in the NIR. Note that even within
individual small clouds $\beta$ can vary significantly. In
Table\,\ref{beta_distribution} we list peak positions and width of the
distribution of $\beta$ for a selection of dark clouds. Most regions belong to
Category 1 or 2, although a small number (e.g Cepheus) can be classified as
Category 3. Note that $\beta$ varies significantly and ranges from 1.0 (e.g.
S\,140) to 2.0 (e.g. Circinus).

\begin{table}
\centering
\renewcommand{\tabcolsep}{4pt}
\caption{\label{beta_distribution} Peak position and width of the distribution
of the opacity index $\beta$ for selected dark clouds, determined between J and
H ($\beta_{\rm JH}$) and between H and K ($\beta_{\rm HK}$). In some regions
$\beta_{\rm HK}$ could not be determined properly, due to low S/N.}
\begin{tabular}{lll|lll}
Region & $\beta_{\rm JH}$ & $\beta_{\rm HK}$ & Region & $\beta_{\rm JH}$ &
$\beta_{\rm HK}$ \\
\noalign{\smallskip}
\hline                                             
\noalign{\smallskip}
\object{CrA} 	   & 1.4$\pm$0.3 & 1.4$\pm$0.4 & \object{LupusIII}  & 1.5$\pm$0.3 & 1.9$\pm$0.4 \\
\object{Ophiuchus} & 1.5$\pm$0.3 & 1.7$\pm$0.4 & \object{LupusIV}   & 1.5$\pm$0.3 & 1.9$\pm$0.3 \\
\object{LupusI}    & 1.5$\pm$0.3 & 1.7$\pm$0.5 & \object{Monocerus} & 1.4$\pm$0.5 & 1.0$\pm$0.4 \\
\object{LupusII}   & 1.3$\pm$0.2 & -           & \object{Cepheus}   & 1.4$\pm$0.4 & 1.2$\pm$0.4 \\
\object{Musca} 	   & 1.7$\pm$0.3 & -           & \object{S140}      & 1.0$\pm$0.3 & 1.5$\pm$0.4 \\ 
\object{Circinus}  & 1.7$\pm$0.4 & 2.0$\pm$0.4 & \object{Serpens}   & 1.7$\pm$0.4 & 2.0$\pm$0.3 \\
\end{tabular}
\end{table}

\section{Conclusions}

We have shown that accumulated star counts in the NIR can be used to obtain
large scale relative extinction maps. Parallel techniques allowed us to process
14400 square degrees of the sky within 40 hours computing time on a 32-node 
Beowulf-type cluster of 2.67\,GHz P4 HyperThreat processors. The accumulated
star counts are performed independent of stellar colour and only local
extinction enhancements are determined, Hence the wavelength dependence of the
dust extinction within the dark clouds can be investigated. A study of selected
clouds shows that the opacity index $\beta$ ranges from 1--2. In most cases a
constant or increasing opacity index with wavelength is found. The large
scatter of $\beta$, even within a particular cloud, leads to the conclusion
that a uniform opacity index should be used with great care for extinction
corrections within dark clouds. This is in contrast to the general ISM where
mixing of the dust, plus averaging over long  lines of sight, ensures a single
index suffices.

\begin{acknowledgements}

We are grateful to the DIAS Cluster Manager, D.~Golden, for scheduling  the
almost 50,000 jobs needed to generate our maps and for his assistance  when a
crucial hard disk failed. D.\,Froebrich and G.C.\,Murphy received support from
the Cosmo-Grid project, funded by the Program for Research in Third Level
Institutions under the National Development Plan and with assistance from the
European Regional Development Fund. The work of A.\,Scholz was partially funded
by Deutsche Forschungsgemeinschaft (DFG) grants Ei409/11-1 and 11-2. This
publication makes use of data products from the Two Micron All Sky Survey,
which is a joint project of the University of Massachusetts and the Infrared
Processing and Analysis Center/California Institute of Technology, funded by
the National Aeronautics and Space Administration and the National Science
Foundation.

\end{acknowledgements}

\end{document}